# Trust-MARL: Trust-Based Multi-Agent Reinforcement Learning Framework for Cooperative On-Ramp Merging Control in Heterogeneous Traffic Flow

Jie Pan[a], Tianyi Wang[b], Christian Claudel[b], Jing Shi[a,†]

[a]*Department of Civil Engineering, Tsinghua University, Beijing, China*

[b]*Department of Civil, Architectural, and Environmental Engineering, The University of Texas at Austin, Austin, Texas, USA*

## Abstract

Intelligent transportation systems require connected and automated vehicles (CAVs) to conduct safe and efficient cooperation with human-driven vehicles (HVs) in complex real-world traffic environments. However, the inherent unpredictability of human behaviour, especially at bottlenecks such as highway on-ramp merging areas, often disrupts traffic flow and compromises system performance. To address the challenge of cooperative on-ramp merging in heterogeneous traffic environments, this study proposes a trust-based multi-agent reinforcement learning (Trust-MARL) framework. At the macro level, Trust-MARL enhances global traffic efficiency by leveraging inter-agent trust to improve bottleneck throughput and mitigate traffic shockwave through emergent group-level coordination. At the micro level, a dynamic trust mechanism is designed to enable CAVs to adjust their cooperative strategies in response to real-time behaviors and historical interactions with both HVs and other CAVs. Furthermore, a trust-triggered game-theoretic decision-making module is integrated to guide each CAV in adapting its cooperation factor and executing context-aware lane-changing decisions under safety, comfort, and efficiency constraints. An extensive set of ablation studies and comparative experiments validates the effectiveness of the proposed Trust-MARL

*[∗,†]Corresponding Author*

*Email addresses:* `panj21@mails.tsinghua.edu.cn` (Jie Pan[a]), `bonny.wang@utexas.edu` (Tianyi Wang[b]), `christian.claudel@utexas.edu` (Christian Claudel[b]), `jingshi@tsinghua.edu.cn` (Jing Shi[a,†])

approach, demonstrating significant improvements in safety, efficiency, comfort, and adaptability across varying CAV penetration rates and traffic densities.

## 1. Introduction

With the rapid development of connected and automated vehicles (CAVs), cooperative control based on CAVs has emerged as a solution to transform urban mobility and intelligent transportation systems (ITS) [1, 2]. However, in heterogeneous traffic environments, where CAVs share roads with human-driven vehicles (HVs), achieving safe, comfortable, and efficient interactions remains challenging [3, 4]. Unlike CAVs, which rely on pre-programmed algorithms to make decisions, HVs exhibit highly variable behaviors, influenced by individual behavioral tendencies, real-time decision-making, and varying levels of attention [5]. Therefore, ensuring CAVs to navigate safely and efficiently among HVs while maintaining ride comfort is crucial to the real-world application of CAVs in ITS [6, 7].

Trajectory planning for CAVs in heterogeneous traffic flow has traditionally relied on game-theoretic methods to model interactions between CAVs and HVs [8]. However, pure game theory often suffer from high computational complexity and limited adaptability, making them less effective in handling HVs' diverse and unpredictable driving styles in real-world scenarios [9]. In particular, a key limitation of these models is their inability to dynamically interpret and adapt to HV behavioral tendencies, leading to suboptimal lane changes, or sudden brakes, which not only increase the risk of accidents, but also disrupt normal traffic flow [10].

To address this gap, multi-agent reinforcement learning (MARL) has emerged as a promising approach for handling complex interactions in heterogeneous traffic [11]. Traditional MARL methods, such as independent Q-learning and actor-critic approaches, have been widely employed to manage interactions between CAVs and HVs [12, 13]. However, these methods often rely on static reward structures and assume fully cooperative environments, limiting their adaptability in heterogeneous traffic scenarios [14]. To solve this limitation, several works have introduced advanced MARL frameworks that encourage agents' cooperation, such as coordinated deep Q-learning and multi-agent actor-critic frameworks [15]. These methods show potential in improving traffic efficiency and safety by opti-mizing CAVs' cooperation. Nonetheless, they typically lack a structured mechanism for incorporating trust in multi-agent interactions, an essential factor for CAVs navigating alongside HVs with unpredictable driving behaviors [16, 17, 18].

Trust, as observed in human interactions, is a dynamic construct shaped by past interactions, behavioral expectations, and real-time observations. It plays a crucial role in enabling cooperative behavior, particularly in scenarios that involve multi-agent coordination under uncertainty [12]. In human-centered autonomous systems, trust has been widely studied as a means to enhance collaboration and mitigate conflict in shared environments [19, 20]. Rather than being an isolated factor, trust can be viewed as a mechanism for realizing cooperation, balancing the agent's own individual objectives with the inferred intentions and behaviors of others. This trust-cooperation coupling becomes especially critical in heterogeneous traffic environments, where agents differ in capabilities and behavioral models. However, existing approaches often treat trust as a static or linearly evolving variable [21], and are largely confined to homogeneous agent settings [22], limiting their generalization and adaptability to real-world traffic scenarios.

To address these limitations, a novel trust-based multi-agent reinforcement learning (Trust-MARL) framework is proposed in this paper, as shown in Figure 1. The core idea of our framework is to integrate both local-level and group-level behaviors through a dynamic trust mechanism that enables CAVs to continuously assess the reliability and behavioral tendencies of neighboring vehicles based on historical interactions and real-time observations. By incorporating this trust-based approach, CAVs can anticipate HV behaviors more accurately and adjust their cooperative strategies accordingly, resulting in safer, more comfort and more efficient interactions in heterogeneous traffic environments.

The main contributions of this paper are outlined as follows:

- **Trust-MARL Framework:** A Trust-MARL framework is proposed, in which trust is modeled as a dynamic, behavior-sensitive variable, enabling adaptive cooperation between CAVs and HVs. Unlike static rule-based or prestructured interaction models, trust in this method is continuously evolved based on historical interactions and real-time observations to modulate cooperative behaviors.

- **Trust-Aware Game-Theoretic Decision-Making:** A trust-aware game-theoretic decision-making model is designed, where cooperative or non-cooperative game is driven by the evolving trust levels. This enables CAVs

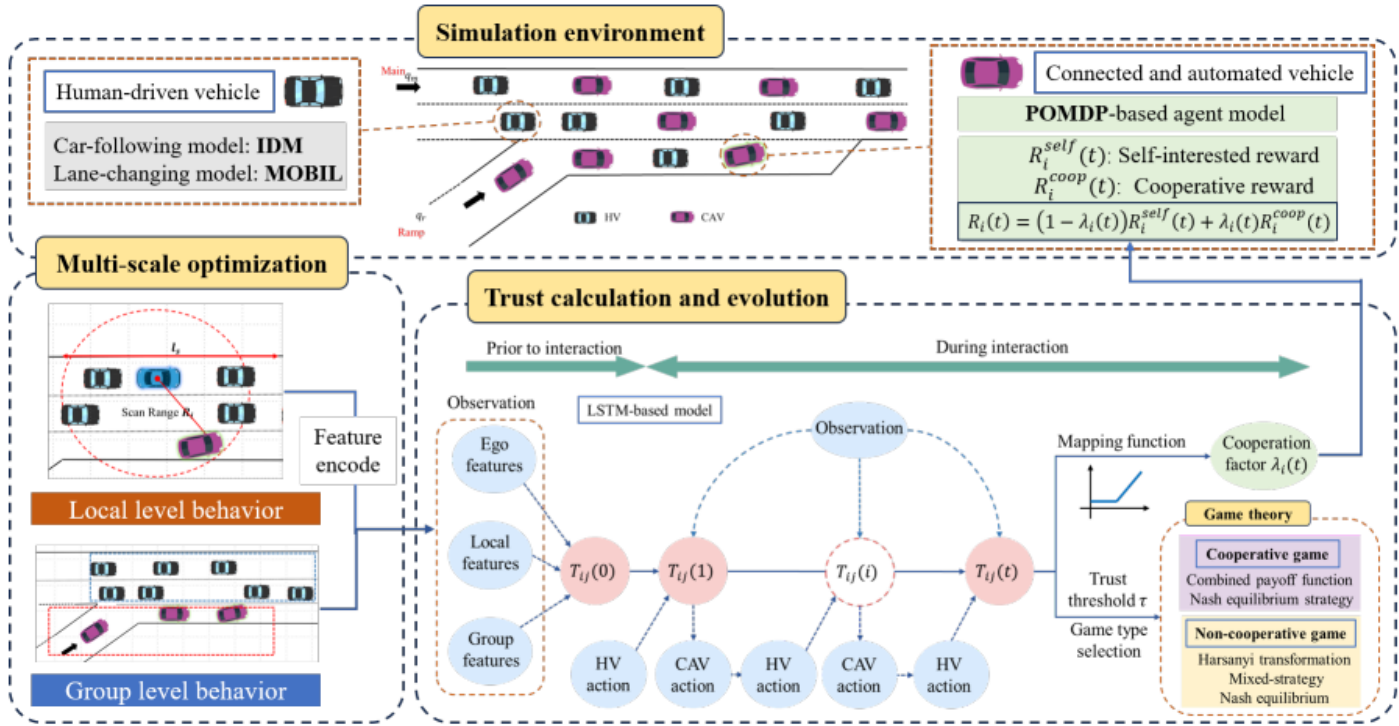

Figure 1: Overview of the Proposed Trust-MARL Framework

to reason about interaction outcomes and adapt its strategy for safety, comfort, and efficiency.

- **Multi-Scale Behavioral Optimization:** Our framework bridges local-level interactions with group-level traffic behaviors. Through trust-sensitive cooperation policies and reward shaping in MARL, the system promotes group-level emergent coordination patterns and macro-level traffic efficiency, enabling robust scalability in heterogeneous traffic environments.

The remainder of this paper is organized as follows: In Section 2, the detailed literature review is conducted. In Section 3, the heterogeneous traffic flow modeling and specific reward function design are introduced. In Section 4, an observation encoding pipeline, a dynamic trust mechanism and a trust-aware game-theoretic lane-changing decision-making model are designed. In Section 5, numerical experiments across diverse traffic conditions are conducted. Experimental results are further discussed in Section 6. Section 7 summarizes the work and discusses future research directions.

## 2. Related Work

*2.1. Multi-Agent Reinforcement Learning in Cooperative On-Ramp Merging*

The focus of MARL-based cooperative on-ramp merging in autonomous driving has shifted to optimizing both microscopic and macroscopic performance metrics. Chen et al. [23] proposed a deep MARL framework with parameter-sharing and well-designed reward functions to improve safety (time to collision (TTC)) and efficiency (average speed) for highway on-ramp merging control. Wu et al. [24] introduced the iPLAN, an intent-aware planning framework using distributed MARL, which enabled CAVs to anticipate and respond to other vehicles' intentions, resulting in improved average jerk and courtesy, and smoother merging operations. Zhang et al. [11] introduced the CCMA, integrating MARL and large language models to foster quicker, smoother, and more adaptive collaboration among CAVs, leading to improved merging success rate, average merging time, collision rate, and traffic throughput. Wang et al. [25] developed several cooperative and non-cooperative MARL algorithms and emphasized that the strategy selection should consider specific traffic scenarios and objectives. Liu et al. [26] developed the GCAV-CPO, which increased efficiency and safety, as well as reduced fuel consumption under different traffic modes and CAV penetration rates.

While MARL enhances decision-making adaptability, it often lacks interpretability and stability. Hybrid approaches combining model-based control with MARL have been proposed to balance efficiency and robustness [27]. Sun et al. [28] introduced a MARL framework integrating model predictive control (MPC) for highway traffic control, effectively handling high-density traffic. Lu et al. [29] incorporated the tube-based enhanced MPC to ensure the safe execution of the MARL policy under disturbances, thereby improving the robustness of traffic control. Li et al. [30] developed a MARL framework combining Bayesian game to derive the optimal cooperative on-ramp merging strategy under diverse driving styles. Lin et al. [31] utilized potential games for modeling and employing deep MARL algorithms for addressing the cooperative on-ramp merging control issue.

*2.2. Game-Theoretic Decision-Making Approaches in Heterogeneous Traffic Flow*

Game theory provides a principle framework for modeling strategic interactions between CAVs and HVs, particularly in heterogeneous traffic environments where uncertainty and asymmetric rationality are prevalent [32]. By modeling agents as utility-driven decision-makers, game-theoretic methods enable CAVs to anticipate and respond to the behaviors of surrounding vehicles under various assumptions of cooperation and competition. Initial studies commonly employed non-cooperative game models to represent individualistic decision-making, particularly in lane-changing and intersection scenarios [33]. Stackelberg games have been widely adopted to model leader-follower dynamics, where CAVs often assume leadership roles while anticipating the reactions of HVs [34]. These models formalized hierarchical interactions and have shown effectiveness in structured scenarios such as ramp merging areas and roundabouts. To reflect bounded human reasoning more accurately, cognitive hierarchy models like Level-$k$ games have been proposed [35, 36, 3]. These frameworks represented different reasoning depths, allowing CAVs to account for heterogeneous cognitive capabilities of human drivers [37]. Such models have been particularly effective in modeling multi-vehicle negotiation at intersections or dense highway scenarios. Recent efforts have extended classical game-theoretic approaches by integrating them with MARL [38]. Markov games and inverse MARL techniques are used to infer human intent and optimize CAV policies under uncertainty [39]. Despite their theoretical strengths, game-theoretic models face challenges in scalability and human behavior modeling, particularly due to the complexity of multi-agent interactions and limited behavioral data.

2.3. Trust Mechanism in Autonomous Driving

Trust has emerged as a critical factor in autonomous driving, particularly in interactions between HVs and CAVs. In this paper, trust refers to the belief in the reliability and cooperative behavior of other vehicles, which guides how CAVs adapt their strategies in uncertain environments. Unlike static assumptions, trust is known to evolve dynamically based on historical interactions, behavioral observations, and contextual understanding [40, 41]. Early research has examined how trust influences driver acceptance, takeover readiness, and interactions [42]. To model trust evolution, probabilistic models such as dynamic Bayesian networks [43] and Gaussian processes [44] have been used to infer and predict trust changes under different behavioral patterns. Recent work has begun incorporating trust into decision-making frameworks for multi-agent systems [45]. For example, trust has been used to modulate the degree of cooperation among agents in MARL, allowing agents to respond adaptively to varying social behaviors [46, 21]. However, these studies often assume homogeneous agent populations and treat trust as a passive signal rather than an active mechanism guiding interaction strategies.

In this work, we view trust as a key to regulating cooperation between agents. By dynamically updating trust levels based on observed behaviors, our Trust-MARL framework enables CAVs to adjust the degree of cooperative behavior during lane-changing and merging process. This allows agents to balance individual objectives with joint traffic outcomes, facilitating both localized interaction and emergent group-level coordination in heterogeneous traffic environments.

## 3. Problem Formulation

We consider a heterogeneous traffic scenario involving HVs and CAVs in an on-ramp merging area on a multi-lane highway. The goal is to develop adaptive lane-changing and car-following policies for CAVs to promote safe, comfortable, efficient, and cooperative interactions with HVs and CAVs.

### 3.1. Heterogeneous Traffic Flow Modeling

The mixed traffic is modeled as a multi-agent environment, where each vehicle is treated as an agent. CAVs act as learning agents governed by the Trust-MARL framework, while HVs follow a rule-based model.

#### 3.1.1. Human-Driven Vehicle

For HVs' car-following behavior, we adopt the intelligent driver model (IDM) [47], which computes longitudinal acceleration based on velocity and distance to the preceding vehicle:

$$a_s(t) = a \left[ 1 - \left( \frac{v_s(t)}{v_d} \right)^4 - \left( \frac{\Delta d^*(t)}{\Delta d_s(t)} \right)^2 \right], \quad (1)$$

$$\Delta d^*(t) = s_0 + \max \left( v_s(t) T + \frac{v_s(t) \cdot \Delta v_s(t)}{2\sqrt{ab}}, 0 \right), \quad (2)$$

where $v_s(t)$ and $\Delta v_s(t)$ denote the current speed and speed difference with the lead vehicle; $\Delta d_s(t)$ and $\Delta d^*(t)$ denote the actual gap and the expected safety distance; and $a$, $b$, $T$, and $s_0$ are model parameters. The specific meaning can be referred to [47].

For lane-changing behavior, we use the minimizing overall braking induced by lane changes (MOBIL) model [48]. A lane change is triggered only if the following two conditions are met:

- **Incentive Criterion:** The current lane change is beneficial to the vehicle, providing acceleration gain while minimizing the impact on other vehicles:

$$\Delta a_{\text{HV}} + p(\Delta a_{\text{f,new}} + \Delta a_{\text{f,old}}) > \Delta a_{\text{th}} \qquad (3)$$

where $\Delta a_{\text{HV}}$ denotes the acceleration increase obtained by the ego vehicle after lane change; $\Delta a_{\text{f,new}} + \Delta a_{\text{f,old}}$ denote the acceleration change by other vehicles; $p$ is a politeness factor used to measure whether HV consider other vehicles; and $\Delta a_{\text{th}}$ is a threshold to avoid frequent lane changes due to minor benefits.

- **Safety Criterion:** Make sure that lane changes do not cause the following vehicle in the new lane to slow down too severely to ensure traffic safety:

$$a_{\text{f,new}} > -b_{\text{safe}} \qquad (4)$$

where $-b_{\text{safe}}$ denotes the maximum acceptable deceleration.

These models jointly reproduce realistic longitudinal and lateral dynamics for HVs in heterogeneous traffic environments.

### 3.1.2. Connected and Automated Vehicle

CAVs are modeled as decentralized agents in a partially observable Markov decision process (POMDP), defined by the tuple:

$$\mathcal{M} = \langle \mathcal{N}, \mathcal{S}, \{\mathcal{A}_i\}, \mathcal{P}, \{\mathcal{R}_i\}, \{\mathcal{O}_i\}, \gamma \rangle \qquad (5)$$

where:

- $\mathcal{N}$: The set of agents.

- $\mathcal{S}$: The global environment state.

- $\mathcal{A}_i$: The joint action space of agent $i$ at each timestep is represented as $\mathbf{a}_t = (a_{\text{long}}, a_{\text{lat}})$: (i) The lateral action space is selected from a discrete set of high-level maneuvers: $a_{\text{lat}} = \{\text{KeepLane, ChangeLeft, ChangeRight}\}$. (ii) The longitudinal action space is modeled as a continuous scalar $a_{\text{long}} \in [-a_{\max}, a_{\max}]$ (the desired acceleration applied at each time step).

- $\mathcal{P}$: The transition function governed by car-following and lane-changing dynamics.

- $\mathscr{R}_i$: The reward function for agent $i$, which is introduced in detail in the next subsection.
- $\mathscr{O}_i$: The observation space of agent $i$, which includes local features within its scan range.
- $\gamma$: The discount factor for future rewards.

Each CAV agent $i \in \mathscr{N}_{\text{CAV}}$ receives an observation $o_i \in \mathscr{O}_i$ of the environment and selects an action $a_i \in \mathscr{A}_i$ based on a policy $\pi_i(o_i)$ to maximize expected discounted reward:

$$\mathscr{J}_i(\pi_i) = \mathbb{E}\left[\sum_{t=0}^{\infty} \gamma^t \mathscr{R}_i(s_t, a_t)\right] \qquad (6)$$

The environment dynamics $\mathscr{P}$ during the training and execution process incorporate both the rule-based reactions of HVs and the POMDP-based interaction decisions of CAVs.

### 3.2. Reward Function Design

Trust is introduced as a latent variable $T_{ij}(t)$, representing agent $i$'s belief in agent $j$'s cooperative intention at time $t$. This trust level modulates cooperation strategies through a weighting factor $\lambda_i(t)$:

$$\mathscr{R}_i(t) = (1 - \lambda_i(t))\mathscr{R}_i^{\text{self}}(t) + \lambda_i(t)\mathscr{R}_i^{\text{coop}}(t) \qquad (7)$$

*Self-Interested Reward.* The self-interested reward $\mathscr{R}_i^{\text{self}}(t)$ for agent $i$ is composed of three weighted components:

$$\mathscr{R}_i^{\text{self}}(t) = w_s \cdot r_i^{\text{safety}}(t) + w_c \cdot r_i^{\text{comfort}}(t) + w_e \cdot r_i^{\text{efficiency}}(t) \qquad (8)$$

with:

$$r_i^{\text{safety}}(t) = -\mathbb{1}\left[d_i^{\text{headway}}(t) < d_{\text{safe}}\right] \qquad (9)$$

$$r_i^{\text{comfort}}(t) = -\left|\frac{da_i(t)}{dt}\right| \qquad (10)$$

$$r_i^{\text{efficiency}}(t) = -|v_i(t) - v_{\text{desired}}| \qquad (11)$$

where:

- $d_i^{headway}(t)$ is the distance to the front vehicle,
- $d_{\text{safe}}$ is the safety threshold,
- $\frac{da_i(t)}{dt}$ is the jerk (first derivative of acceleration),
- $v_i(t)$ is the current speed,
- $v_{\text{desired}}$ is the desired free-flow speed,
- $\mathbb{1}[x]$ is defined as a random variable which is 1 if x is true, and 0 if x is not true.

*Cooperative Reward.* The cooperative reward $\mathscr{R}_i^{\text{coop}}(t)$ is computed as a trust-weighted average of the self-rewards of nearby agents:

$$\mathscr{R}_i^{\text{coop}}(t) = \frac{1}{Z_i(t)} \sum_{j \in \mathscr{N}_i(t)} T_{ij}(t) \cdot \mathscr{R}_j^{\text{self}}(t) \tag{12}$$

where $Z_i(t) = \sum_{j \in \mathscr{N}_i(t)} T_{ij}(t)$ is a normalization factor, and $\mathscr{N}_i(t)$ denotes the set of neighbors within the scan range.

*Cooperation Factor.* To integrate trust into the agent's utility optimization, we define a trust-based cooperation factor $\lambda_i(t)$, which determines the balance between self-interested and cooperative rewards in the total reward signal:

$$\lambda_i(t) = \min\left(1, \max\left(0, \frac{T_{ij}(t) - \tau}{1 - \tau}\right)\right), \tag{13}$$

where $\tau$ is a cooperation threshold. A higher trust value increases the influence of cooperative components in the reward function, encouraging socially aligned lane-changing and car-following behavior.

## 4. Methodology

### 4.1. Observation Encoding and Temporal Modeling

To enable context-aware decision-making in partially observable and dynamic traffic environments, we design an observation encoding pipeline that captures both spatial and temporal aspects of traffic scenes. Each CAV agent encodes its surroundings via structured inputs and a long short-term memory (LSTM)-based sequential model to learn behavior-consistent representations.

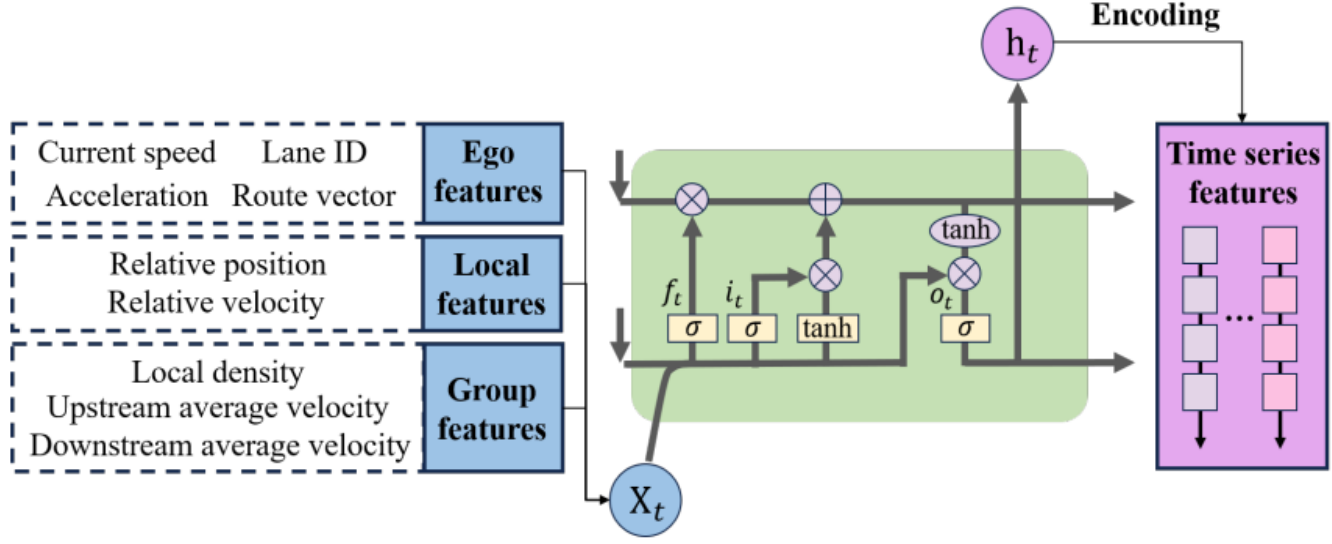

Figure 2: Architecture of Observation Encoding and LSTM-Based Temporal Modeling

### 4.1.1. Multi-Scale Observation Space

The raw observation vector for agent $i$ at time $t$, denoted $\mathbf{o}_i^t$, is composed of three feature subsets:

- **Ego Features** $o_i^{\text{ego}}$: current speed $v_i^t$, acceleration $a_i^t$, lane ID $l_i^t$, route vector $r_i^t$.

- **Local Features** $o_i^{\text{local}}$: relative positions and velocities $[\Delta x_{ij}^t, \Delta v_{ij}^t]$ of surrounding vehicles $j \in \mathcal{N}_i^t$ within a scan range $d_s$.

- **Group Features** $o_i^{\text{group}}$: flow-level indicators, including local density $\rho_i^t$ and average velocity $\bar{v}_i^t$ in the upstream/downstream group of $i$.

The full observation vector is given by:

$$\mathbf{o}_i^t = [o_i^{\text{ego}}, o_i^{\text{local}}, o_i^{\text{group}}] \tag{14}$$

### 4.1.2. Temporal Encoding with LSTM

To model historical dependencies and behavioral trends, each agent maintains a sliding observation history $\mathbf{O}_i^t = [\mathbf{o}_i^{t-k}, \ldots, \mathbf{o}_i^t]$ across $k$ time steps. The temporal encoding is performed as:

$$h_i^t = LSTM_\theta(\mathbf{O}_i^t) = f_{\text{LSTM}}(\mathbf{o}_i^{t-k:t}; \theta) \tag{15}$$

where $\theta$ denotes trainable LSTM parameters; and $h_i^t$ is the hidden representation that summarizes temporal interaction features.

This encoded representation $h_i^t$ is subsequently used as input to the policy learning network $\pi$:

$$a_i^t \sim \pi(a|h_i^t) \tag{16}$$

As shown in Figure 2, the encoded input consists of ego and neighbor state features across time, with spatial dependencies captured by the scan range and temporal dynamics learned by the LSTM backbone.

### 4.2. Dynamic Trust Mechanism

Trust serves as a dynamic representation of social belief between agents, guiding whether a CAV should cooperate with another agent in heterogeneous traffic environments. We introduce a dynamic trust mechanism that enables each CAV to continuously evaluate, update, and apply trust levels toward its neighboring vehicles, forming the foundation for trust-aware decision-making and reward adaptation.

#### 4.2.1. Trust Definition and Update Rule

Let $T_{ij}(t) \in [0,1]$ denote the trust level that agent $i$ assigns to agent $j$ at time step $t$. Trust is dynamically updated after each interaction using an exponentially weighted formulation:

$$T_{ij}(t+1) = (1-\hat{\alpha})T_{ij}(t) + \hat{\alpha} \cdot \delta(a_j^t), \tag{17}$$

where $\hat{\alpha} \in (0,1)$ is a smoothing factor and $\delta(a_j^t) \in \{0,1\}$ indicates whether agent $j$'s behavior at time $t$ is perceived as cooperative from agent $i$'s perspective. Specifically:

$$\delta(a_j^t) = \begin{cases} 1, & \text{if } a_j^t \text{ contributes to agent } i \text{ achieving its goal,} \\ 0, & \text{otherwise.} \end{cases} \tag{18}$$

This update mechanism allows the trust signal to encode both short-term behavior and long-term cooperative trends, dynamically reflecting the reliability and social alignment of neighboring agents.

*4.2.2. Trust-Driven Interaction Adjustment*

The learned trust also informs policy-level decision-making. For instance, when $T_{ij}(t) < \tau$, the agent may opt for a conservative response (e.g., defensive lane-keeping), while $T_{ij}(t) \geq \tau$ triggers cooperative maneuvers such as yielding or lane-gap creation. This trust-calibrated behavior promotes emergent cooperation in heterogeneous traffic flow, without requiring explicit communication. By embedding trust into both reward shaping and policy modulation, our dynamic trust mechanism provides an interpretable, scalable, and data-driven way to manage interactions between heterogeneous agents in decentralized multi-agent settings.

*4.3. Trust-Aware Game-Theoretic Decision-Making Process*

In this section, we formulate a trust-aware game-theoretic decision-making process that guides the selection of cooperative or non-cooperative policies for each CAV agent using trust dynamics. This integration enables agents to adaptively modulate their behavior in heterogeneous traffic scenarios based on evolving trust levels, balancing local-level incentives with group-level coordination.

*4.3.1. Trust-Guided Game Type Determination*

We model each interaction between a CAV and its surrounding vehicles (CAVs or HVs) as a repeated two-player game:

$$\text{GameType}_{ij}^t = \begin{cases} \text{Cooperative,} & \text{if } T_{ij}(t) \geq \tau, \\ \text{Non-Cooperative,} & \text{otherwise.} \end{cases} \quad (19)$$

This classification governs the structure of the agent's payoff function and subsequent policy optimization.

- **Cooperative Game:** In this study, a non-zero-sum game can be only initiated by a CAV. A non-zero-sum game refers to a game in which all players receive a payoff corresponding to their actions and the sum of their payoffs is not zero. Once the CAV receives a reply to the cooperation invitation from other CAV, a two-player non-zero-sum cooperative game will be formed.

- **Non-Cooperative Game:** If the CAV receives no response from the other participants, a two-player non-zero-sum non-cooperative game under incomplete information will be formed. In this case, the Harsanyi transformation is adopted, which transforms a game of incomplete information to a game of imperfect information.

A key difference between cooperative and non-cooperative game is that in cooperative games, players can make binding agreements before playing the game, e.g., how to share payoffs, while agreements are not binding in non-cooperative games. On the other hand, the individual players are simply unaware of actions chosen by the other players in non-cooperative games while cooperative games consider collaboration of players.

4.3.2. Cooperative Game

In this case, the players can make binding agreements before playing the game, e.g., how to share payoffs $R_{ij}(S_i, S_j)$. Later, a probable set of actions in the game and a comprehensive set of payoff functions addressing traffic efficiency, ride comfort, and lane-changing efficiency will be introduced. The payoff matrices for the ego vehicle and the new follower are shown in Table 1, mainly inherent from SECRM-2D [49]. The combined payoff functions for vehicles in the cooperative game is given by:

$$U_{coop}(S_i, S_j) = R_{ij}(S_i, S_j) = \gamma_{eff} \cdot R_{eff}(S_i, S_j) \\ + \gamma_{comf} \cdot R_{comf}(S_i, S_j) \\ + \gamma_{lc} \cdot R_{lc}(S_i, S_j) \\ + \gamma_{mlc} \cdot R_{mlc}(S_i, S_j)$$

(20)

where $R_{eff}, R_{comf}, R_{lc},$ and $R_{mlc}$ are the payoffs of efficiency, comfort, mandatory and discretionary lane-changing efficiency, respectively; and $\gamma_{eff}, \gamma_{comf}, \gamma_{lc},$ and $\gamma_{mlc} \geq 0$ are the corresponding weights.

The Nash equilibrium strategy can be formulated as:

$$S_i^* = \arg\max_{S_i} U_{coop}(S_i, S_j).$$

(21)

Table 1: Payoff Matrices in Cooperative and Non-Cooperative Games

| New Follower's Strategy | Cooperative (Payoffs) | | Non-Cooperative (Payoffs) | | Non-Cooperative (Probability) | |
|---|---|---|---|---|---|---|
| | Change Lane | Not Change Lane | Change Lane | Not Change Lane | Change Lane | Not Change Lane |
| Accelerate | $R_{11}$ | $R_{12}$ | $(P_{11}, Q_{11})$ | $(P_{12}, Q_{12})$ | $p_1$ | $q_1$ |
| Keep | $R_{21}$ | $R_{22}$ | $(P_{21}, Q_{21})$ | $(P_{22}, Q_{22})$ | $p_2$ | $q_2$ |
| Decelerate | $R_{31}$ | $R_{32}$ | $(P_{31}, Q_{31})$ | $(P_{32}, Q_{32})$ | $p_3$ | $q_3$ |
| Change Lane | $R_{41}$ | $R_{42}$ | $(P_{41}, Q_{41})$ | $(P_{42}, Q_{42})$ | $p_4$ | $q_4$ |

### 4.3.3. Non-Cooperative Game

In this case, the agreements are not binding and the individual players are the cornerstone in the non-cooperative game. After applying Stackelberg game principles and Harsanyi transformation, we can compute the payoff functions across various action combinations, aiming to identify an optimal action for both the ego vehicle and the new follower. In this paper, $p_n$ and $q_n$ represent probabilities of different actions taken by the new follower when the ego vehicle changes lane and not changes lane, respectively. They satisfy $p_n > 0$, $q_n > 0$, $\sum_{n=1,2,3,4} p_n = 1$ and $\sum_{n=1,2,3,4} q_n = 1$. In the human driving phase, this probability distribution reflects human driver's driving style; while in the fully automatic driving phase, it is a design parameter of the self-driving system. Building on the previous study CLACD [50], the payoff matrices for the ego vehicle and the new follower are shown in Table 1. The payoff functions for vehicles in the non-cooperative game is given by:

$$\begin{cases} P_{ij}(S_i) = \alpha_{ij}^0 + \alpha_{ij}^1 Acc_{ij}(S_i) + \alpha_{ij}^2 \Delta V(S_i) + \varepsilon_{ij}, \\ Q_{ij}(S_j) = \beta_{ij}^0 + \beta_{ij}^1 Acc_{ij}(S_j) + \beta_{ij}^2 \Delta V(S_j) + \delta_{ij}. \end{cases} \quad (22)$$

where $P$ and $Q$ denote the payoffs for the ego vehicle and new follower, respectively; $Acc$ represents acceleration; $\Delta V$ is speed change; $\varepsilon$ and $\delta$ represent the error term; and $\alpha$ and $\beta$ are parameters to be estimated.

According to the idea of mixed-strategy Nash equilibrium, the ego vehicle takes the maximization of the expectation of its own payoff functions as the optimal strategy, and the combination of the strategy space and the mixed-strategy probability distribution of the new follower can be used to compute the expected payoff functions under different strategies of the ego vehicle. The combined payoff functions in the non-cooperative game is given by:

$$U_{non-coop}(S_i, S_j) = \begin{cases} \sum_{n=1,2,3,4}(P_{n1} + p_n Q_{n1}) \\ \sum_{n=1,2,3,4}(P_{n2} + q_n Q_{n2}) \end{cases} \quad (23)$$

## 5. Experiments

### 5.1. Experimental Setup

We conduct experiments using the simulation of urban mobility (SUMO) [51] to evaluate our proposed method in a highway on-ramp merging scenario. The simulation environment replicates a 10-kilometer, two-lane highway with a dedicated on-ramp section shown in Figure 3, where CAVs and HVs interact under varying traffic demands. The mainline lanes are 3.75 m wide and limited to 33.3 m/s. The ramp carries an advisory speed of 22.2 m/s before vehicles adapt to mainline flow. Induction-loop detectors are placed every 250 m on each lane and conflict-point sensors in the merging zone capture flow, speed, occupancy, minimum TTC, and post-encroachment time. HVs follow the IDM car-following model and MOBIL lane-changing model whereas CAVs are governed by our Trust-MARL controller that exchanges state information within a 100 m radius at 10 Hz. The highway includes a merging region where vehicles entering from the on-ramp must integrate into the main traffic stream. The merging region is designed to test cooperative behavior, lane-changing decision-making, and the impact of different CAV penetration rates. Key parameters for simulation setup are summarized in Table 2. Four CAV penetration rates (0 %, 30 %, 70 %, and 100 %) are tested under three demand levels (300, 600, and 900 veh/h) , with arrivals split 80% mainline and 20% ramp to maintain consistent merging pressure:

- **Traffic Compositions:** CAVs and HVs are mixed in varying proportions: 0% CAVs and 100% HVs, 30% CAVs and 70% HVs, 70% CAVs and 30% HVs, and 100% CAVs and 0% HVs.

- **Traffic Densities:** Low (300 vehicles/hour), Medium (600 vehicles/hour), and High (900 vehicles/hour).

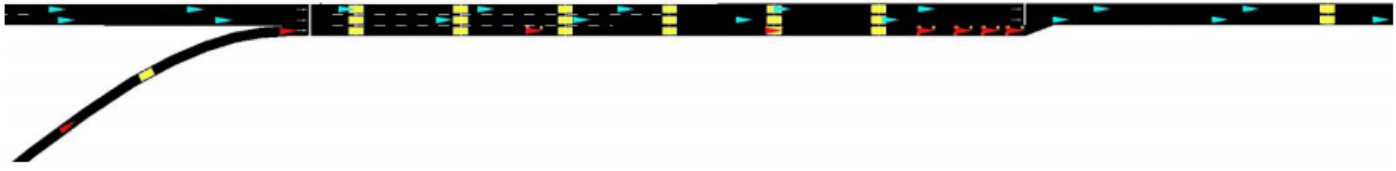

Figure 3: Highway On-Ramp Merging Scenario in SUMO

Each factor combination runs for one hour of warm-up followed by five hour of data collection, and ten random seeds introduce stochastic variability in driver traits. Collected data support assessment of efficiency (travel time), safety (conflicts), comfort (longitudinal jerk), and adaptability (recovery time), providing a statistically robust basis for quantifying the benefits of the proposed trust-aware cooperative merging framework across congestion states and autonomy levels.

Table 2: Simulation and Model Parameters

| Parameter | Value |
|---|---|
| *Simulation Setup* | |
| Simulation duration | 18,000 seconds (5 hours) |
| Highway length | 10 km |
| Number of main lanes | 2 |
| Number of ramp lanes | 1 |
| Maximum speed (main/ramp) | 33.3 / 22.2 m/s |
| Replay buffer size | 100,000 |
| Batch size | 64 |
| Scan range $d_s$ | 100 m |
| *Trust Model* | |
| Initial trust $T_{ij}(0)$ | 0.5 |
| Smoothing factor $\hat{\alpha}$ | 0.6 |
| Trust level threshold $\tau$ | 0.4 |
| *POMDP and Reward Function Parameters* | |
| Discount factor $\gamma$ | 0.95 |
| Safety threshold $d_{safe}$ | 4 m |
| Free-flow desired speed $v_{desired}$ | 30 m/s |
| Weights: $w_s$, $w_e$, $w_c$ | 1.0 |
| *IDM (Car-Following)* | |
| Maximum acceleration $a$ | 1.0 m/s$^2$ |
| Comfortable deceleration $b$ | 1.5 m/s$^2$ |
| Desired time headway $T$ | 1.2 s |
| Minimum spacing $s_0$ | 2.0 m |
| Desired speed $v_d$ | 30 m/s |
| *MOBIL (Lane-Changing)* | |
| Politeness factor $p$ | 0.2 |
| Advantage threshold $\Delta a_{th}$ | 0.3 m/s$^2$ |
| Safe braking limit $b_{safe}$ | 4.0 m/s$^2$ |
| *Game-Theoretic Utility Weights* | |
| $\gamma_{eff}$, $\gamma_{comf}$ | 0.3, 0.3 |
| $\gamma_{lc}$, $\gamma_{mlc}$ | 0.2, 0.2 |

## 5.2. Training Setup and Comparison

To evaluate the effectiveness of our proposed Trust-MARL framework, we conduct experiments under a structured curriculum learning regime in a SUMO-based highway on-ramp merging environment. The setup simulates realistic traffic conditions by varying both demand levels and vehicle compositions. To facilitate stable convergence and enhance generalization, we adopt a structured curriculum learning strategy that progressively increases environmental complexity throughout training. The curriculum begins with a low-density setting (300 vehicles/hour) under a fully autonomous scenario (100% CAV), enabling agents to acquire fundamental car-following and merging behaviors in an ideal cooperative context. In the next stage, we introduce partial autonomy (e.g., 70% CAV and 30% HV) under the same low traffic density, allowing agents to adapt to heterogeneous interactions and initiate trust modeling. As training advances, vehicle density is increased to 600 vehicles/hour with a 50% CAV composition, challenging the scalability of the learned policies under moderate interaction complexity. The fourth stage simulates congested scenarios (900 vehicles/hour) with varied CAV penetration rates (30% and 70%), requiring robust trust-aware coordination strategies to sustain safe and efficient merging. Finally, a randomized replay stage is introduced, where traffic densities and vehicle compositions are dynamically varied across episodes. This final phase ensures that agents generalize effectively to unseen traffic configurations, enhancing robustness and adaptability prior to deployment.

We compare our Trust-MARL against four widely-used MARL algorithms:

- **MADDPG** [52]: Multi-agent deterministic policy gradient with a centralized critic.
- **MAPPO** [53]: Decentralized actor with centralized proximal policy optimization (PPO)-based updates.
- **MAA2C** [54]: Advantage actor-critic adapted for multi-agent cooperation.
- **MASAC** [55]: Entropy-regularized actor-critic that encourages diverse exploration.

All baselines use the same reward structure and observation design for comparison. Our Trust-MARL augments MASAC with a trust-aware cooperation module. Figure 4 presents the training performance of four baseline MARL algorithms—MADDPG, MAPPO, MAA2C, and MASAC—compared against our proposed Trust-MARL framework. Results are averaged over 10 runs with 90% confidence intervals. To ensure a fair and consistent comparison, we adopt a unified experimental setting across all models, focusing solely on varying the underlying training paradigms while holding environmental factors and reward structures constant. Rather than contrasting models with and without trust, we instead evaluate diverse learning strategies to isolate the impact of training mechanisms.

Among the baseline methods, MASAC shows the strongest performance due to its entropy-driven exploration, which encourages policy diversity and robust-ness. Building upon this foundation, Trust-MARL introduces trust-based policy shaping, allowing agents to selectively engage in cooperation based on evolving trust levels. This result confirms that integrating trust dynamics into the training process not only boosts the overall reward but also improves convergence and stability. Consequently, MASAC is selected as the base learner for our Trust-MARL framework due to its favorable balance between exploration and learning stability. Although every algorithm eventually converges, the Trust-MARL framework consistently outperforms its counterparts by converging more rapidly, attaining higher final episodic rewards through its dynamic trust mechanism that strengthens long-term cooperation and decision quality, and exhibiting lower performance variance across random seeds, which attests to its superior generalization and robustness.

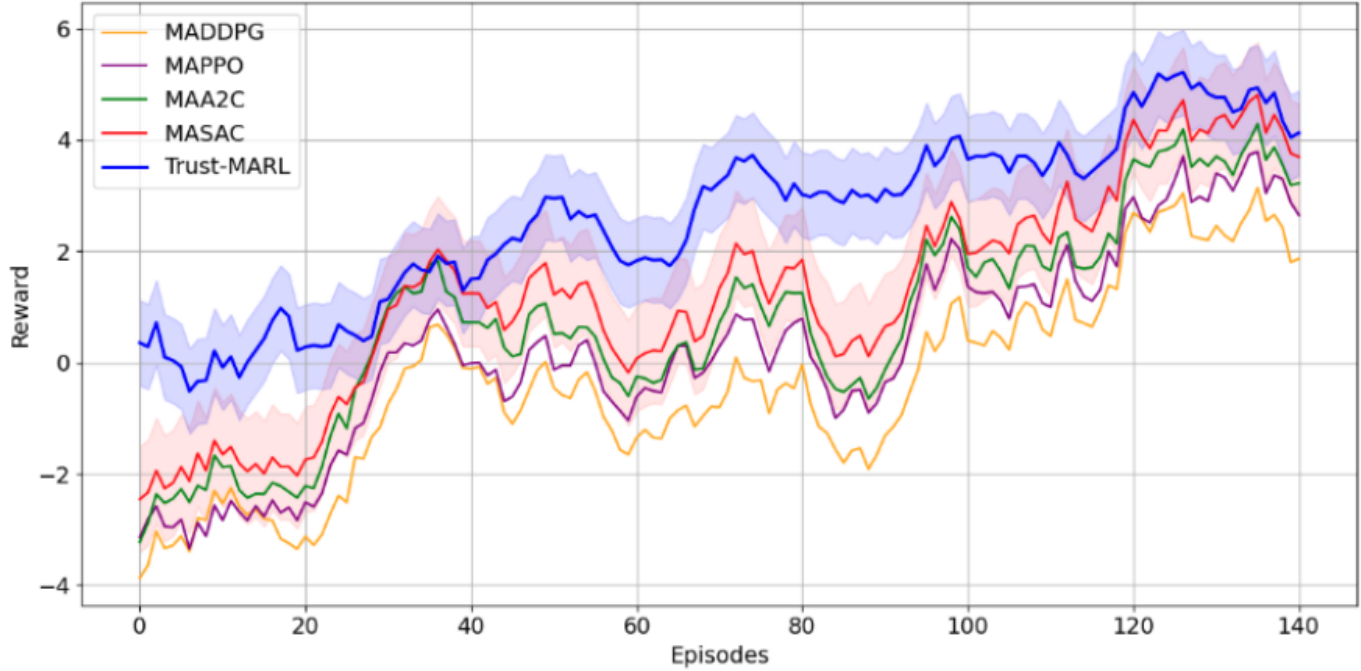

Figure 4: Training Reward Comparison Between our Trust-MARL and Baselines

## 6. Results and Discussion

### 6.1. Ablation Study and Module Contributions

To systematically evaluate the contributions of each core module in our framework—namely, (1) dynamic trust mechanism, (2) game-theoretic decision-making, and (3) MARL, we design an ablation study with four controlled configurations. Each variant selectively removes one or more components to isolate their respective impacts on safety, efficiency, and comfort.

The study evaluates four configurations for multi-agent autonomous driving,

Table 3: Ablation Study in Trust-MARL Framework

| Method | MARL | Trust | Game Theory | Collision Rate Red. (%) | Travel Time Red. (%) | Jerk Red. (%) |
|---|---|---|---|---|---|---|
| Game Theory | ✗ | ✗ | ✓ | 2.1 | 6.8 | 13.2 |
| MARL | ✓ | ✗ | ✗ | 1.7 | 5.4 | 11.0 |
| MARL + Game Theory | ✓ | ✗ | ✓ | 1.5 | 4.9 | 9.8 |
| **Trust-MARL (Ours)** | ✓ | ✓ | ✓ | **1.2** | **3.8** | **7.4** |

Note: Reductions are relative to a Rule-Based baseline [56] under high-density traffic conditions.

and the findings in Table 3 confirm that fully integrating the dynamic trust mechanism, game-theoretic reasoning, and MASAC learning yields the strongest gains in safety, efficiency, and comfort. The Game Theory setup, relying on cooperative and non-cooperative game strategies without learning based on [57], succeeds in eliciting some cooperative merges but lacks on-line adaptation. As a result, it reacts sluggishly to unforeseen driver behaviour, leading to elevated collision and jerk levels. By contrast, the MARL-only approach quickly tailors its policy to local traffic but, in dense flow, tends to overfit to short-term rewards and occasionally triggers aggressive maneuvers that undermine coordination robustness. Combining Game Theory with MARL partially remedies this, because explicit role definitions accelerate convergence toward socially acceptable lane-changing protocols. Nevertheless, without a trust filter, the agents are prone to over-commit to cooperation and brake harshly when partners defect, explaining their only moderate improvements. The Trust-MARL (Full) model, resolves these shortcomings: the trust layer continuously recalibrates each agent's willingness to yield or merge, preventing abrupt braking when a neighbouring vehicle shows erratic intent, while MASAC fine-tunes low-level control and the game-theoretic module preserves global reciprocity. Consequently, Trust-MARL cuts collisions to 1.2 %, trims mean travel time by 3.8 %, and lowers longitudinal jerk by 7.4 % relative to the rule-based baseline—benefits that translate into measurable gains in ride comfort and throughput at scale. The proposed Trust-MARL model highlights the strength of combining social reasoning, coordination incentives, and expressive learning policies. This analysis validates the importance of each module and confirms that our Trust-MARL achieves superior performance in heterogeneous traffic environments.

## 6.2. Comparative Experiments

### 6.2.1. Rationale and Description of Comparative Methods

To comprehensively evaluate the proposed Trust-MARL framework for on-ramp merging, we design comparative experiments across three traffic densities and four CAV penetration rates. The experiments aim to quantify improvements in safety, efficiency, and comfort by comparing our approach against state-of-the-art baseline methods, highlighting the benefits of trust-aware cooperative mechanisms in autonomous driving. The methods compared are as follows:

- **Rule-Based** [56]: A baseline method employing fixed rules for on-ramp merging, lacking learning or adaptability, often used in traffic simulations to model traditional vehicle behavior.

- **Non-Coop MARL** [58]: A baseline method utilizing MARL where vehicles learn individual policies without explicit cooperation, reflecting a decentralized but uncoordinated approach.

- **Trust-Sensitive HVs** [59]: Incorporate trust modeling and game-theoretic principles to guide human-like vehicle interactions during merging, but does not leverage MARL for learning.

- **Proposed Trust-MARL**: Our integrated framework combining dynamic trust mechanism, game theory, and MARL (using MASAC), designed to optimize safety, efficiency, and comfort through trust-aware coordination.

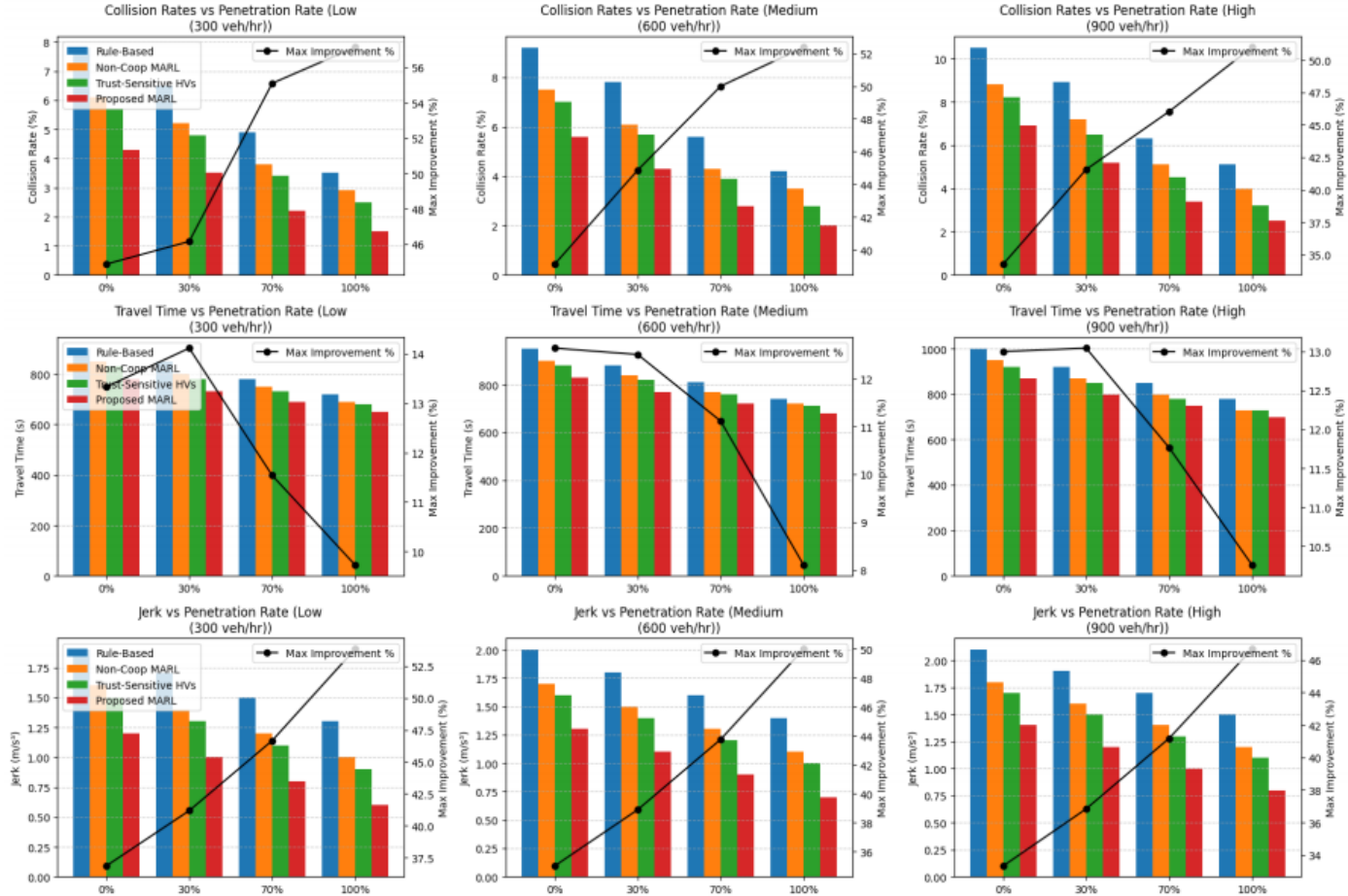

Figure 5: Comparison of Collision Rates, Travel Time, and Jerk across Traffic Conditions

Figure 5 systematically evaluates the collision rates, travel times, and jerk values across three traffic density levels, low (300 veh/hr), medium (600 veh/hr), and high (900 veh/hr), under varying CAV penetration rates (0%, 30%, 70%, and 100%) for four distinct methods: Rule-Based, Non-Coop MARL, Trust-Sensitive HVs, and the Proposed Trust-MARL. The analysis reveals consistent and significant enhancements in safety, efficiency, and comfort resulting from the implementation of the Proposed Trust-MARL framework.

### 6.2.2. Safety Analysis - Collision Rates

As shown in Figure 5 (top row), collision risk falls monotonically as the proportion of CAVs grows, yet the magnitude of improvement is highly method-dependent. In the Rule-Based baseline, collision rates remain stubbornly high. With purely human-driven traffic at 900 veh/hr, 10.5 % of merges still result in conflict, and even at full CAV penetration rates, the collision rate only drops to 5.2 %, reflecting the limited conflict-resolution capability of fixed heuristics. The Non-Coop MARL agents learn to exploit small gaps and thus reduce collisions more effectively at low and medium densities, but the absence of explicit coordination causes policy oscillations in dense flow, leaving a 4 % collision rate at full CAVs. Trust-Sensitive HVs, which embed game-theoretic reasoning but forego learning, exhibit safer performance than Non-Coop MARL in heavy traffic because their reciprocity rules encourage early yielding, yet their static payoff structure cannot capitalize on the additional maneuverability offered by higher CAV shares, so collision rates plateau around 3.2 %.

The Proposed Trust-MARL architecture delivers the clearest safety gains. By fusing game-theoretic intent inference with a dynamic trust , CAVs temper aggressive merges when nearby vehicles act unpredictably, while MASAC optimization continuously refines gap-creation tactics. Consequently, collision rates fall below 3 % once CAV penetration rate exceeds 70 % and reach 2.5 % at full CAVs under the most demanding scenario, a 51 % reduction relative to the Rule-Based benchmark. Notably, Trust-MARL also indicates greater robustness to random driver heterogeneity. The results show that pure cooperation without adaptive trust makes agents over-commit or under-coordinate, while learning alone is brittle in dense traffic unless anchored by social priors. By fusing trust estimation, game-theoretic reasoning, and MARL, Trust-MARL meets both needs and sharply reduces lane-change conflicts, enabling safer CAV operation on congested highways.

### 6.2.3. Efficiency Analysis - Travel Time

The middle row of Figure 5 presents average travel time across varying traffic densities. It shows a clear drop in average travel time as the share of CAVs rises, yet only the Proposed Trust-MARL fully realizes this potential across all densities. When the traffic density is low, the gap between different methods is small. At medium density, Rule-Based vehicles begin to queue behind hesitant mergers, pushing mean travel time to about 800 s, Trust-MARL's advantage expands. Notably, in high-density scenarios at full CAV penetration rates, the Proposed Trust-MARL method reduces travel time from 780 seconds (Rule-Based) to 700 seconds, achieving a 10% improvement. The Trust-Sensitive HVs method demonstrates modest travel time reductions, yet its incomplete cooperative nature limits its performance. Although the Non-Coop MARL outperforms the Rule-Based approach, it fails to reach optimal performance due to inadequate coordination during vehicle merging processes. These observations indicate that trust-based cooperation significantly enhances traffic flow efficiency by facilitating smoother on-ramp merging behavior and reducing congestion-induced delays.

### 6.2.4. Comfort Analysis - Jerk Values

Figure 5 (bottom row) highlights jerk values as indicators of ride comfort. Under high-density conditions with full CAV penetration rates (100%), the Proposed Trust-MARL achieves the lowest jerk value of 0.8 m/s$^3$, while the Rule-Based method results in a value of 1.5 m/s$^3$. This corresponds to a 46.7% reduction in jerk, indicating significantly smoother driving behavior. Trust-Sensitive HVs and Non-Coop MARL methods yield intermediate improvements but fall short of the comfort performance demonstrated by the Proposed Trust-MARL. The consistently low jerk values achieved by the Proposed Trust-MARL underscore the critical role of cooperative, trust-aware decision-making in enhancing ride comfort and vehicle stability. Moreover, smoother longitudinal profiles translate directly into reduced occupant fatigue and lower mechanical stress on drivetrain components, suggesting tangible maintenance and health benefits for large-scale CAV fleets.

### 6.2.5. Adaptability Analysis - Recovery Time

To quantitatively assess adaptability, we introduce the concept of *Recovery Time*. This metric captures how quickly the traffic system returns to a stable performance regime after a sudden reduction in CAV penetration rates (e.g., from 70% to 30%). Specifically, we define recovery time as the number of simulation steps required for the mainline throughput to return to at least 90% of its preshift value. This reflects the system's ability to adapt to heterogeneous traffic transitions in real time. Table 4 summarizes the adaptability performance of all the aforementioned methods. It can be seen that the Rule-Based method cannot converge within 1000 seconds while our Trust-MARL approach only needs 410 seconds to recover convergence.

Table 4: Comparison of recovery time

| Method | Adaptability (Recovery Time) |
|---|---|
| Rule-Based | >1000s |
| Non-Coop MARL | 820s |
| Trust-Sensitive HVs | 610s |
| **Trust-MARL (Ours)** | **410s** |

### 6.3. Traffic Flow Analysis under Dynamic Demand

To evaluate the robustness and coordination effectiveness of the proposed Trust-MARL framework under real-world-like conditions, we simulate a dynamic traffic flow scenario with time-varying demand. The demand profile, shown in Figure 6, features a peak inflow rate of 1000 vehicles/hour on the mainline and 250 vehicles/hour on the ramp. This time-varying setup creates realistic congestion phases, allowing for thorough comparison of merging and cooperation performance across different learning-based and rule-based strategies. Figure 7 reports the resulting *bottleneck flow*, *ramp inflow*, and *mainstream throughput* over the 6-hour simulation. It demonstrates that the proposed Trust-Based MARL (black curve) sustains the greatest throughput across all three subsystems. At the bottleneck, it consistently delivers 10–15 % more vehicles per hour than the next-best controller during the first 90 minutes and postpones the inevitable post-1.5 h capacity drop by roughly 15 minutes, evidencing more disciplined gap-creation and fewer disruptive slowdowns. On the ramp, absolute volumes are modest but Trust-MARL still records a slightly higher, markedly smoother profile, indicating steadier platoon release onto the mainline. In the mainstream corridor, Trust-MARL sustains an average flow of roughly 1050 veh/hr, whereas the baselines remain in the 900–1000 veh/hr range. In addition, Trust-MARL sees the collapse toward zero flow roughly 15 minutes earlier than the baselines, indicating that its cooperative strategy exhausts the available headway sooner while the baselines maintain moderate throughput a bit longer. Overall, Figure 7 highlights the framework's capacity to maximize throughput and prolong stable operation by harmonizing merging decisions across ramp and mainline traffic.

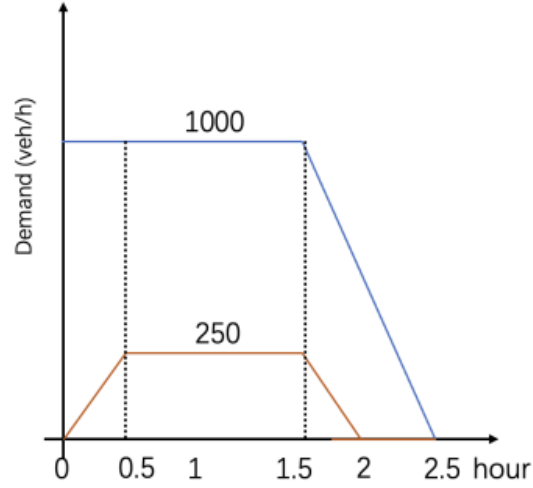

Figure 6: Dynamic Demand Profile for Highway Mainline and Ramp

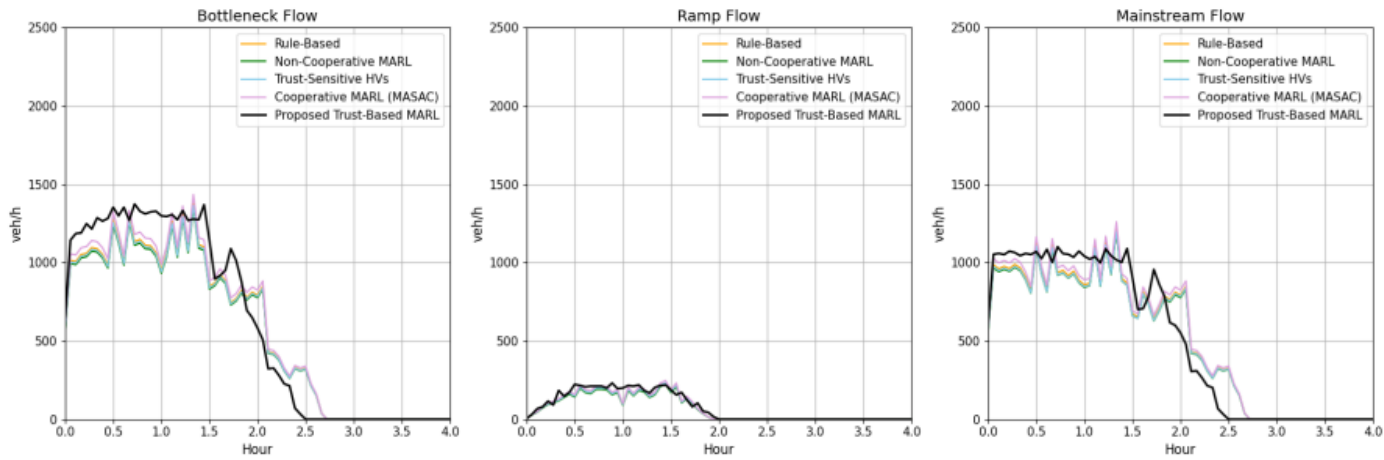

Figure 7: Traffic Flow Comparison across Baselines under Dynamic Demand: (Left) Bottleneck Flow, (Middle) Ramp Inflow, (Right) Mainline Throughput

Our Trust-MARL consistently achieves superior flow stability and throughput across all traffic zones. Specifically, it sustains higher ramp inflow and smoother mainline merging under congested conditions, reflecting the benefits of adaptive cooperation.

## 7. Conclusion

This work presents a comprehensive framework for trust-based multi-agent reinforcement learning (Trust-MARL) tailored for autonomous driving in heterogeneous traffic environments. By unifying trust evolution, game-theoretic lane-changing decision-making, and cooperative reward shaping, we introduce a dynamic trust mechanism that enables connected and autonomous vehicles (CAVs) to dynamically adjust their behavior based on trust levels toward surrounding agents. The proposed Trust-MARL captures both local-level vehicle interactions and group-level traffic dynamics, offering a scalable solution for real-time decision-making in on-ramp merging scenarios. Our theoretical formulation and experimental evaluations collectively demonstrate that incorporating trust and game theory into MARL leads to significant improvements in safety, efficiency, comfort, and adaptability. Empirical results reveal that our Trust-MARL outperforms rule-based baselines and conventional MARL algorithms across all key metrics, and effectively maintains stable throughput under varying CAV penetration rates and traffic density levels. These outcomes validate our core hypothesis that trust-aware coordination enables more socially intelligent and context-sensitive traffic behaviors.

Despite these promising results, several research challenges remain. First, *robustness against adversarial behavior*—such as deceptive agents exploiting the trust mechanism—requires further investigation through adversarial training or robust estimation techniques. Second, enhancing *generalization across heterogeneous scenarios*, including stochastic human driver models and diverse road geometries, is vital to ensure wide applicability. Third, improving the *explainability of trust modeling and policy decisions* will be critical for user trust and regulatory acceptance. Fourth, we advocate for *multi-scale trust adaptation*, where trust is modeled not only between individual vehicles but also among groups or infrastructure-level entities. Lastly, future work may benefit from interdisciplinary approaches that integrate social cognition, human factors, and behavioral economics to further refine agent cooperation and enhance human-likeness in heterogeneous traffic flow. In summary, this study provides a foundational step toward the integration of social constructs—such as trust—into autonomous decision-making, paving the way for safer, more adaptive, and socially compliant intelligent transportation systems.

## 8. Acknowledgment

## References

[1] Y. Wang, T. Wang, Research on dual clutch intelligent vehicle infrastructure cooperative control based on system delay prediction of two lane highway

on ramp merging area, Automotive Innovation 7 (2024) 588–601. doi:https://doi.org/10.1007/s42154-024-00283-2.

[2] T. Wang, Q. Guo, C. He, H. Li, Y. Xu, Y. Wang, J. Jiao, Impact of connected and automated vehicles on longitudinal and lateral performance of heterogeneous traffic flow in shared autonomy on two-lane highways, SAE Technical Paper (2025-01-8098) (2025). doi:https://doi.org/10.4271/2025-01-8098.

[3] S. Fang, P. Hang, C. Wei, Y. Xing, J. Sun, Cooperative driving of connected autonomous vehicles in heterogeneous mixed traffic: A game theoretic approach, IEEE Transactions on Intelligent Vehicles (2024) 1–15doi:10.1109/TIV.2024.3399694.

[4] Y. Li, X. Wang, T. Wang, Q. Liu, Characteristics analysis of autonomous vehicle pre-crash scenarios, arXiv preprint arXiv:2502.20789 (2025). doi:https://doi.org/10.48550/arXiv.2502.20789.

[5] X. Zhao, Z. Wang, Z. Xu, Y. Wang, X. Li, X. Qu, Field experiments on longitudinal characteristics of human driver behavior following an autonomous vehicle, Transportation Research Part C: Emerging Technologies 114 (2020) 205–224. doi:https://doi.org/10.1016/j.trc.2020.02.018.

[6] Y. Wang, Q. Liu, Z. Jiang, T. Wang, J. Jiao, H. Chu, B. Gao, H. Chen, Rad: Retrieval-augmented decision-making of meta-actions with vision-language models in autonomous driving, arXiv preprint arXiv:2503.13861 (2025). doi:https://doi.org/10.48550/arXiv.2503.13861.

[7] Y. Wang, Z. Huang, Q. Liu, Y. Zheng, J. Hong, J. Chen, L. Xiong, B. Gao, H. Chen, Drive as veteran: Fine-tuning of an onboard large language model for highway autonomous driving, in: 2024 IEEE Intelligent Vehicles Symposium (IV), IEEE, 2024, pp. 502–508. doi:10.1109/IV55156.2024.10588851.

[8] D. Wu, Z. Li, C. Du, C. Liu, Y. Li, X. Xu, Lane changing control of autonomous vehicle with integrated trajectory planning based on stackelberg game, IEEE Open Journal of Intelligent Transportation Systems 5 (2024) 810–825. doi:10.1109/OJITS.2024.3509462.

[9] V. G. Lopez, F. L. Lewis, M. Liu, Y. Wan, S. Nageshrao, D. Filev, Game-theoretic lane-changing decision making and payoff learning for autonomous vehicles, IEEE Transactions on Vehicular Technology 71 (4) (2022) 3609–3620. doi:10.1109/TVT.2022.3148972.

[10] C. Xu, W. Zhao, C. Wang, T. Cui, C. Lv, Driving behavior modeling and characteristic learning for human-like decision-making in highway, IEEE Transactions on Intelligent Vehicles 8 (2) (2023) 1994–2005. doi:10.1109/TIV.2022.3224912.

[11] M. Zhang, Z. Fang, T. Wang, S. Lu, X. Wang, T. Shi, Ccma: A framework for cascading cooperative multi-agent in autonomous driving merging using large language models, Expert Systems with Applications 282 (2025) 127717. doi:https://doi.org/10.1016/j.eswa.2025.127717.

[12] P. Hansen-Estruch, I. Kostrikov, M. Janner, J. G. Kuba, S. Levine, Idql: Implicit q-learning as an actor-critic method with diffusion policies, arXiv preprint arXiv:2304.10573 (2023). doi:https://doi.org/10.48550/arXiv.2304.10573.

[13] Y. Wang, D. Shi, C. Xue, H. Jiang, G. Wang, P. Gong, Ahac: Actor hierarchical attention critic for multi-agent reinforcement learning, in: 2020 IEEE International Conference on Systems, Man, and Cybernetics (SMC), 2020, pp. 3013–3020. doi:10.1109/SMC42975.2020.9283339.

[14] S. Aradi, Survey of deep reinforcement learning for motion planning of autonomous vehicles, IEEE Transactions on Intelligent Transportation Systems 23 (2) (2022) 740–759. doi:10.1109/TITS.2020.3024655.

[15] T. Shi, P. Wang, X. Cheng, C.-Y. Chan, D. Huang, Driving decision and control for automated lane change behavior based on deep reinforcement learning, in: 2019 IEEE Intelligent Transportation Systems Conference (ITSC), 2019, pp. 2895–2900. doi:10.1109/ITSC.2019.8917392.

[16] J. Haoran, Y. Yuyu, H. Qiang, Z. Pengqian, G. Ting, Multi-agent trust evaluation model based on reinforcement learning, in: 2021 8th International Conference on Dependable Systems and Their Applications (DSA), 2021, pp. 608–613. doi:10.1109/DSA52907.2021.00088.

[17] J. Dinneweth, A. Boubezoul, R. Mandiau, S. Espié, Multi-agent reinforcement learning for autonomous vehicles: a survey, Autonomous Intelligent Systems 2 (1) (2022) 27. doi:https://doi.org/10.1007/s43684-022-00045-z.

[18] W. Wang, F. Hui, J. Zhang, S. Fang, Deep reinforcement learning method for trajectory planning of connected and autonomous vehicles in the roundabout lane-changing scenario, in: 2024 4th International Symposium on Computer Technology and Information Science (ISCTIS), 2024, pp. 168–173. doi:10.1109/ISCTIS63324.2024.10698974.

[19] X. Wang, Z. Shi, F. Zhang, Y. Wang, Mutual trust based scheduling for (semi)autonomous multi-agent systems, in: 2015 American Control Conference (ACC), 2015, pp. 459–464. doi:10.1109/ACC.2015.7170778.

[20] C. Basu, M. Singhal, Trust dynamics in human autonomous vehicle interaction: A review of trust models, in: Proceedings of the 2016 AAAI Spring Symposium Series, 2016.

[21] Z. Zhou, G. Liu, Y. Tang, Multiagent reinforcement learning: Methods, trustworthiness, applications in intelligent vehicles, and challenges, IEEE Transactions on Intelligent Vehicles (2024) 1–23doi:10.1109/TIV.2024.3408257.

[22] J. Monteil, R. Billot, J. Sau, F. Armetta, S. Hassas, N. E. El Faouzi, Cooperative highway traffic: multi-agent modelling and robustness to local perturbations, in: Transportation Research Board 92nd Annual Meeting, 2013, p. 24 p.

[23] D. Chen, M. Hajidavalloo, Z. Li, K. Chen, Y. Wang, L. Jiang, Y. Wang, Deep multi-agent reinforcement learning for highway on-ramp merging in mixed traffic, IEEE Transactions on Intelligent Transportation Systems 24 (11) (2023) 11623–11638. doi:10.1109/TITS.2023.3285442.

[24] X. Wu, R. Chandra, T. Guan, A. Bedi, D. Manocha, iplan: Intent-aware planning in heterogeneous traffic via distributed multi-agent reinforcement learning, in: Proceedings of The 7th Conference on Robot Learning (CoRL), Vol. 229, 2023, pp. 446–477.
URL https://proceedings.mlr.press/v229/wu23b.html

[25] D. Wang, G. Xu, Z. Wu, To cooperate or not: Multi-agent reinforcement learning-based on-ramp merging strategies for autonomous vehicles, in: 2024 IEEE 20th International Conference on Automation Science and Engineering (CASE), 2024, pp. 2415–2420. doi:10.1109/CASE59546.2024.10711529.

[26] L. Liu, X. Li, Y. Li, J. Li, Z. Liu, Reinforcement-learning-based multilane cooperative control for on-ramp merging in mixed-autonomy traffic, IEEE Internet of Things Journal 11 (24) (2024) 39809–39819. doi:10.1109/JIOT.2024.3447039.

[27] W. Li, F. Qiu, L. Li, Y. Zhang, K. Wang, Simulation of vehicle interaction behavior in merging scenarios: A deep maximum entropy-inverse reinforcement learning method combined with game theory, IEEE Transactions on Intelligent Vehicles 9 (1) (2024) 1079–1093. doi:10.1109/TIV.2023.3323138.

[28] D. Sun, A. Jamshidnejad, B. D. Schutter, A novel framework combining mpc and deep reinforcement learning with application to freeway traffic control, IEEE Transactions on Intelligent Transportation Systems 25 (7) (2024) 6756–6769. doi:10.1109/TITS.2023.3342651.

[29] K. Lu, D. Li, Q. Wang, K. Yang, L. Zhao, Z. Song, Safe reinforcement learning-based eco-driving control for mixed traffic flows with disturbances, IEEE Transactions on Intelligent Transportation Systems (2025) 1–12 doi:10.1109/TITS.2025.3544812.

[30] L. Li, W. Zhao, C. Wang, Cooperative merging strategy considering stochastic driving style at on-ramps: A bayesian game approach, Automotive Innovation 7 (2024) 312–334. doi:https://doi.org/10.1007/s42154-023-00248-x.

[31] Q. Lin, W. Huang, Z. Wu, M. Zhang, Z. He, Multi-agent game theory-based coordinated ramp metering method for urban expressways with multi-bottleneck, IEEE Transactions on Intelligent Transportation Systems 26 (3) (2025) 3643–3658. doi:10.1109/TITS.2024.3521460.

[32] X. Li, W. Zhang, H. Chen, Survey on game-theoretic multi-vehicle interaction decision-making for intelligent driving, Control Theory & Applications 40 (2023) 1–15. doi:http://dx.doi.org/10.12345/kzyjc.20230502.

[33] J. Wang, M. Li, Game analysis of driver behavior in mixed motorized-nonmotorized traffic flow, Journal of Traffic and Transportation Engineering 14 (12) (2014) 89–97. doi:https://doi.org/10.3969/j.issn.1671-1637.2014.12.010.

[34] Q. Zhao, Y. Liu, Multi-vehicle cooperative decision-making in mixed traffic based on stackelberg game, China Journal of Highway and Transport 37 (3) (2024) 45–56. doi:https://doi.org/10.19721/j.cnki.1001-7372.2024.03.004.

[35] Q. Zhang, R. Langari, H. E. Tseng, S. Mohan, S. Szwabowski, D. Filev, Stackelberg differential lane change game based on mpc and inverse mpc, IEEE Transactions on Intelligent Transportation Systems (2024). doi:10.1109/TITS.2024.3386790.

[36] D. Li, H. Pan, Two-lane two-way overtaking decision model with driving style awareness based on a game-theoretic framework, Transportmetrica A: transport science 19 (3) (2023) 2076755. doi:https://doi.org/10.1080/23249935.2022.2076755.

[37] Y. Yan, L. Peng, T. Shen, J. Wang, D. Pi, D. Cao, G. Yin, A multi-vehicle game-theoretic framework for decision making and planning of autonomous vehicles in mixed traffic, IEEE Transactions on Intelligent Vehicles 8 (11) (2023) 4572–4587. doi:10.1109/TIV.2023.3321346.

[38] S. Karimi, A. Karimi, A. Vahidi, Level-$k$ reasoning, deep reinforcement learning, and monte carlo decision process for fast and safe automated lane change and speed management, IEEE Transactions on Intelligent Vehicles 8 (6) (2023) 3556–3571. doi:10.1109/TIV.2023.3265311.

[39] Z. Liu, T. Zhang, Driving style recognition and game-theoretic lane change decision in mixed traffic, Automotive Engineering 45 (9) (2023) 1567–1578. doi:https://doi.org/10.19562/j.chinasae.2023.09.001.

[40] J. Pan, J. Shi, Safety evaluation and prediction of overtaking behaviors in heterogeneous traffic considering dynamic trust and automated driving styles, Transportation Research Part F: Traffic Psychology and Behaviour 109 (2025) 383–398. doi:https://doi.org/10.1016/j.trf.2024.12.020.

[41] P. Yu, S. Dong, S. Sheng, L. Feng, M. Kwiatkowska, Trust-aware motion planning for human-robot collaboration under distribution temporal logic specifications, in: 2024 IEEE International Conference on Robotics and Automation (ICRA), IEEE, 2024, pp. 12949–12955. doi:10.1109/ICRA57147.2024.10610874.

[42] C. Wilson, D. Gyi, A. Morris, R. Bateman, H. Tanaka, Non-driving related tasks and journey types for future autonomous vehicle owners, Transportation Research Part F: Traffic Psychology and Behaviour 85 (2022) 150–160. doi:https://doi.org/10.1016/j.trf.2022.01.004.

[43] A. Xu, G. Dudek, Optimo: Online probabilistic trust inference model for asymmetric human-robot collaborations, in: Proceedings of the 10th Annual ACM/IEEE International Conference on Human-Robot Interaction, 2015, pp. 221–228. doi:https://doi.org/10.1145/2696454.2696492.

[44] H. Soh, Y. Xie, M. Chen, D. Hsu, Multi-task trust transfer for human–robot interaction, The International Journal of Robotics Research 39 (2-3) (2020) 233–249. doi:https://doi.org/10.1177/0278364919866905.

[45] R. Valiente, B. Toghi, M. Razzaghpour, R. Pedarsani, Y. P. Fallah, Learning-based social coordination to improve safety and robustness of cooperative autonomous vehicles in mixed traffic, Machine Learning and Optimization Techniques for Automotive Cyber-Physical Systems (2023) 671–707doi:https://doi.org/10.1007/978-3-031-28016-0_24.

[46] H. L. Fung, V.-A. Darvariu, S. Hailes, M. Musolesi, Trust-based consensus in multi-agent reinforcement learning systems, arXiv preprint arXiv:2205.12880 (2022). doi:https://doi.org/10.48550/arXiv.2205.12880.

[47] M. Treiber, A. Hennecke, D. Helbing, Congested traffic states in empirical observations and microscopic simulations, Physical Review E 62 (2) (2000) 1805–1824. doi:10.1103/PhysRevE.62.1805.

[48] A. Kesting, M. Treiber, D. Helbing, General lane-changing model mobil for car-following models, Transportation Research Record 1999 (1) (2007) 86–94. doi:10.3141/1999-10.

[49] T. Shi, I. Smirnov, O. ElSamadisy, B. Abdulhai, Secrm-2d: Rl-based efficient and comfortable route-following autonomous driving with analytic safety guarantees, arXiv preprint arXiv:2407.16857 (2024). doi:https://doi.org/10.48550/arXiv.2407.16857.

[50] Y. Ali, Z. Zheng, M. M. Haque, M. Yildirimoglu, S. Washington, Clacd: A complete lane-changing decision modeling framework for the connected and traditional environments, Transportation Research Part C: Emerging Technologies 128 (2021) 103162. doi:https://doi.org/10.1016/j.trc.2021.103162.

[51] P. A. Lopez, M. Behrisch, L. Bieker-Walz, J. Erdmann, Y.-P. Flötteröd, R. Hilbrich, Microscopic traffic simulation using sumo, in: 2018 IEEE Intelligent Transportation Systems Conference (ITSC), 2018, pp. 2575–2582. doi:10.1109/ITSC.2018.8569938.

[52] R. Lowe, Y. Wu, A. Tamar, J. Harb, P. Abbeel, I. Mordatch, Multi-agent actor-critic for mixed cooperative-competitive environments, arXiv preprint arXiv:1706.02275 (2017). doi:https://doi.org/10.48550/arXiv.1706.02275.

[53] C. Yu, A. Velu, E. Vinitsky, J. Gao, Y. Wang, A. Bayen, Y. Wu, The surprising effectiveness of ppo in cooperative multi-agent games, Advances in neural information processing systems 35 (2022) 24611–24624.

[54] T. Chu, Y. Ye, Y. Zhan, H. Ma, J. Zhu, Z. Zhang, Multi-agent actor-critic with hierarchical graph attention networks, in: Proceedings of the AAAI Conference on Artificial Intelligence, Vol. 34, 2020, pp. 4296–4303.

[55] A. Touati, H. Satija, J. Romoff, J. Pineau, P. Vincent, Randomized value functions via multiplicative normalizing flows, in: Uncertainty in Artificial Intelligence, PMLR, 2020, pp. 422–432.

[56] Y. Zhang, Y. Zhuang, C. Zhu, A rule-based cooperative merging strategy for connected and automated vehicles, IEEE Transactions on Intelligent Transportation Systems 23 (8) (2022) 11578–11589. doi:10.1109/TITS.2019.2928969.

[57] X. Liao, X. Zhao, Z. Wang, K. Han, P. Tiwari, M. J. Barth, G. Wu, Game theory-based ramp merging for mixed traffic with unity-sumo co-simulation,

IEEE Transactions on Systems, Man, and Cybernetics: Systems 52 (9) (2022) 5746–5757. doi:10.1109/TSMC.2021.3131431.

[58] S. Shalev-Shwartz, S. Shammah, A. Shashua, Safe, multi-agent, reinforcement learning for autonomous driving, arXiv preprint arXiv:1610.03295 (2016). doi:https://doi.org/10.48550/arXiv.1610.03295.

[59] H. S. Ahmad, E. Sabouni, W. Xiao, C. G. Cassandras, W. Li, Trust-aware resilient control and coordination of connected and automated vehicles, in: 2023 IEEE 26th International Conference on Intelligent Transportation Systems (ITSC), IEEE, 2023, pp. 4309–4314. doi:10.1109/ITSC57777.2023.10421858.